\documentclass[graybox, envcountchap]{svmult}

\usepackage[square,sort,comma,numbers]{natbib}
\usepackage{mathptmx}        
\usepackage{amsmath}
\usepackage{amssymb}
\usepackage{color}
\usepackage{helvet}          
\usepackage{courier}         
\usepackage{dirtree}

\usepackage{makeidx}        
\usepackage{graphicx}        
\usepackage{subfig}

\usepackage{multicol}        
\usepackage[bottom]{footmisc}

\usepackage[colorlinks=true,citecolor=blue,linkcolor=blue,urlcolor=blue]{hyperref}

\usepackage[misc]{ifsym}
\usepackage{longtable}

\makeindex             

\begin{document}


\title{Charge exchange in X-ray astrophysics}
\author{Liyi Gu and Chintan Shah}
\institute{Liyi Gu (\Letter) \at SRON Netherlands Institute for Space Research, Niels Bohrweg 4, 2333 CA Leiden, the Netherlands, \email{l.gu@sron.nl}
\and Chintan Shah \at NASA Goddard Space Flight Center, 8800 Greenbelt Rd., Greenbelt, MD 20771, USA; \at Max-Planck-Institut f$\rm \ddot{u}$r Kernphysik, Saupfercheckweg 1, D-69117 Heidelberg, Germany, \email{chintan.shah@mpi-hd.mpg.de}}
%
%
\maketitle

\abstract{Charge exchange is an atomic process that primarily occurs at interfaces between the neutral and ionized gas. The study of the process has been carried out on three levels: the theoretical calculation of the cross sections, the laboratory measurements of reaction rates and line strengths, and the observational constraints using celestial objects. For a long time in the past, the status of astrophysical observations in the X-ray band lagged behind the other two aspects until the discovery of X-ray from a comet was made in 1996, which changed the research landscape. Recent observational evidence suggests that charge exchange has been seen or can be expected from a surprisingly broad range of locations, from the Earth's exosphere to the large-scale structures of the Universe. The rapid development of high-resolution X-ray spectroscopy, in particular the non-dispersive micro-calorimeters, is paving the way to revolutionary new science possibilities both in the laboratory and astrophysics. This chapter summarizes the current knowledge of charge exchange and its relevance on astrophysics, especially X-ray spectroscopy.}


\section{Introduction}
\label{sec:1}


The phenomenon of ion-neutral interaction was studied in early laboratory experiments. The scattering experiment of alpha particles or bare He ions, against gold atoms has led \citet{rutherford1911} to establish the classical model of atomic structure. The quantum mechanical treatment of the ion-neutral scattering problem was explored by \citet{massey1933}, and \citet{mulliken1939}, among others, pioneered the theoretical spectral modeling of charge exchange using atomic orbital approximation. From 1950s, as the gas-beam technique matured, an increasing number of experiments have been put forward to measure directly the charge exchange cross sections \cite{stier1956, fite1958}. Early investigations of this process have been covered in several reviews \citep{gilbody1981, knudsen1982, janev85}.        

It has been well known that the atom-ion charge exchange by far dominates all kinds of electron-ion collisions. For a collision of $\sim$ keV/u, the effective cross section of charge exchange between highly-ionized ions and neutrals are at least two orders of magnitude greater than the radiative recombination and electron-impact excitation cross sections. From the ion view, charge exchange provides efficient recombination on the ionization state. Unlike radiative recombination, charge exchange always ends up in an excited state, producing strong line emissions but zero continuum.

Charge exchange has been well utilised in the research of nuclear fusion. The radiative transitions from excited states of impurity ions resulting from charge exchange may constitute a substantial portion of the total energy losses; meanwhile, the charge exchange emission can be used as a powerful tool for spectroscopic diagnostics
of high-temperature plasmas (e.g., \citealt{car1987}). 

The study of charge exchange has been extended to astrophysics, in particular the interaction between solar wind protons with neutral particles near the Sun, in the planets, and from the interstellar medium (e.g., \citealt{wallis1973}). Charge exchange has also been found to be the origin of the broad H$\alpha$ emission lines observed at the shock wave of supernova remnants \citep{cr1978}. These lines are emitted from hot hydrogen atoms created by electron capture of hot protons in the post-shock plasma.

An important breakthrough was made in the late 1990s, when the first detection of cometary X-rays
\citep{obs1} and the charge exchange interpretation \citep{cravens1997} was made. Since then, the interest on charge exchange X-ray has been activated in observations, but also in theory and in the laboratory. It has been recognized as a primary or second emission mechanism in a broad range of astrophysicobjects, as well as a time-varying background emission tl X-ray sources. In this review, we provide an overview of our current knowledge on charge exchange, with a focus on X-ray astrophysics with spectroscopic utilised In section~\ref{sec:2} we present the basic physical properties of charge exchange, as well as the theoretical and experimental achievements. Section~\ref{sec:3} shows the observational results obtained so far. The review
will close with conclusions and a short outlook in section~\ref{sec:4}.

\section{Plasma modeling}
\label{sec:2}
\subsection{Physics and classic models}
\label{sec:physics}

\begin{figure}[bht]
\sidecaption
\includegraphics[scale=.45]{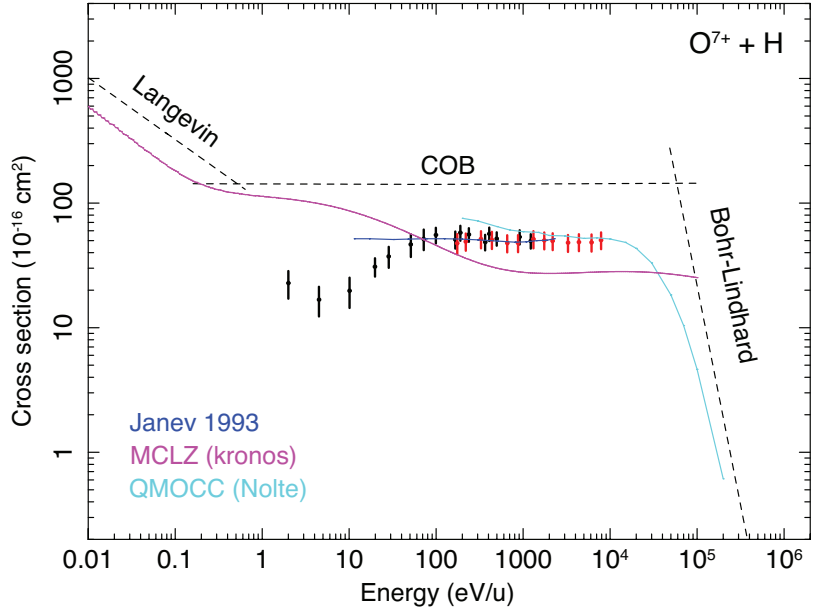}
\caption{Total cross sections for O$^{7+}$ + H based on the experimental measurements of \citet{exp88} (black) and \citet{1979PhRvA..19..515M} (red). The theoretical calculations from \citet{janev1993}, multichannel Landau-Zener \citep{theo40}, and quantum molecular-orbital close-coupling \citep{nolte2012} are shown in blue, purple, and cyan. The classical Langevin, over-the-barrier, and Bohr-Lindhard models are plotted in dashed lines.}
\label{fig:totcs}       
\end{figure}

Charge exchange recombination consists of the transfer of one or more electrons from (mostly) the outer atomic shell of target atom/ion $M$ to the outer shell of an impinging/projectile atom/ion $X$, leading to the recombination of the impinging particle while the ionization of $M$. The process can be written as

\begin{equation}
X^{q+} + M \rightarrow X^{(q-m)+} + M^{m+},    
\end{equation}
where $q$ is the original charge of the $X$, and $m$ is the number of electrons transferred. $m = 1$ can be referred as single electron capture (SEC) from the point of view of the impinging ion. $m = 2$ is designated as double electron capture (DEC).    

The total cross section for this process is typically of order of 10$^{-15}$ - 10$^{-14}$ cm$^{2}$, greater by a factor of $10^{2} - 10^{3}$ than the cross sections of typical electron-impact radiative recombination and dielectronic recombination. The classical analytic calculations, as summarized in \citet{stancil01}, provide order estimate of the total cross sections. They include
\begin{itemize}
    \item Langevin model for $< 0.1$ eV/u collision, where $\rm u$ stands for atomic mass unit. This model predicts an energy dependence $E^{-1/2}$ of the cross section, which is found to be in line with the Landau-Zener approximation shown in Figure~\ref{fig:totcs}.
    \item Classical over-the-barrier (COB) model for intermediate collision speed. In this approximation, the cross section becomes a constant over energy.
    \item Bohr-Lindhard model, which predicts a $E^{-7/2}$ behavior, becomes applicable at very high energy.
\end{itemize}

These classic treatments for $O^{7+} + H$ are shown in Figure~\ref{fig:totcs} with a comparison to various explicit theoretical and experimental results. These simple models seem to reproduce the correct general trend as a function of energy, though apparently overestimating the absolute values. 

Charge exchange operates in a semi-resonant favor when a subset of the energy levels of the projectile ion and the target overlap. If the projectile ion is highly charged, its highly excited levels overlap with the neutral's ground state, making it possible to transfer electrons onto high-$n$ levels. The distribution of $n$ with the most likelihood has been studied in various ways,
\begin{itemize}
\item Simple COB model predicts the dominant $n$ level $n_{\rm max} \leq q / \sqrt{2 I_{M}}$, where $q$ is the initial charge and $I_{M}$ is the ionization potential of the target.
\item Systematic compilation of existing results by \citet{janev85} reported a scaling $n_{\rm max} \sim q (1 + (q - 1) / \sqrt(2 q) )^{-0.5} \sim q^{3/4}$. 
\item More recently, \citet{gu2016a} suggested that the $n$ distribution is energy dependent; the capture is peaked on a single $n$ channel at low collision energy, while it tends to be shared among several neighbour $n$ channels at the intermediate energy range. The analytic form of the transfer is reported as Eq.A.1 of their paper. At the very high energy end, the $n$ distributions become narrow again \citep{wa2008}. 
\end{itemize}
For the He-like triplet, the cascade contribution from large $n$ to the formation of forbidden lines is substantially larger than that to the resonance lines, yielding a large forbidden-to-resonance line ratio.

To construct an emission spectrum from charge exchange, it is necessary to model the population of capture onto states with orbital angular momenta $l$ in the range of (0, $n$-1). For collisions with $E \geq 10$ keV/u, it is expected that the $l$ distribution function $W$ is weighed statistically, $W_{\rm st}$ = $(2 l + 1) / n^{2}$. Therefore the largest possible $l = n - 1$ will be the most populated. As the collision velocity decreases, the capture $l$ tends to decrease towards $l = 1$ or 2 taking one of the following possible forms,
\begin{itemize}
    \item Low energy approximation L1, $W_{\rm L1}$ = (2$l$ + 1) [($n$ - 1)!]$^2$ / [($n$ + $l$)! ($n$ - $l$ - 1)!],
    \item Modified low energy approximation or L2, $W_{\rm L2}$ = $l$ ($l$ + 1) (2$l$ + 1) ($n$ - 1)!($n$ - 2)! / [($n$ + $l$)!($n$ - $l$ - 1)!],
    \item Shifted low energy approximation or L3, $W_{\rm L3}$ = (2$l$ + 3)  [($n$ - 1)!]$^2$ / [($n$ + $l$ + 1)! ($n$ - $l$ - 2)!],
    \item ``Separable'', $W_{\rm se}$ = (2$l$ + 1)/$q$ e$^{-l(l + 1)/q}$,
    \item even or constant over $l$, $W_{\rm ev}$ = 1 / $n$.
\end{itemize}
For a high-speed collision, shells with large $l$ are significantly populated, resulting in more optical and UV lines following the Yrast cascade relaxation. The low-speed charge exchange would instead yield more X-ray lines because the shells with small $l$ become more populated.

Commonly the capture on the total spin quantum number is assumed to be statistical, and therefore the triplet-to-singlet ratio of He-like ion, produced from the H-like ion reaction, is 3. However, bias from the statistical weight is found in recent theoretical calculations (e.g., \citealt{wu2011, nolte2012}). These works suggested that the triplet-to-singlet ratios, and therefore the distribution on the spin quantum number, are likely to be energy-dependent.

Transitions from highly-excited levels with large $n$ and the large forbidden-to-resonance line ratio are two basic spectral features of the charge exchange X-ray emission. The high-$n$ Rydberg lines are usually stronger in H-like than in He-like series because the cascades in the He-like ions are subject to the singlet-to-triplet branching ratios. The UV-to-X-ray ratios of charge exchange emission are larger than those of the electron-impact emission.

\subsection{Theoretical calculations of absolute cross sections}

\onecolumn
\begin{longtable}[H]{ c | c | c | c }
\caption[longtable]{Theoretical data} \\
\label{theo_data} 
\centering
Reference &  type$^{a}$  & ion & energy (eV/u)  \\
\endfirsthead
\centering
\centering
Reference &  type$^{a}$  & ion & energy (eV/u)  \\
\hline
\endhead
\endfoot
\hline
\citet{theo12} & total & $\rm D + H^{+}$, $\rm D^{+} + H$ & $\leq 6$ \\
\citet{theo67} & total,$nl$ & $\rm He^{2+} - Ne^{10+} + H, Li$ & 10$^3$ - 10$^5$\\
\citet{theo4} & total & $\rm Li + H^{+}$ & 10$^{-4}$ - 10  \\
\citet{theo3} & total & $\rm Li + H^{+}$ & 10$^{-5}$ - 10$^3$  \\
\citet{theo20} & total,$nl$ & $\rm A^{A+} + H$ (5 $\leq A \leq 74$) & 30 - 8$\times 10^4$ \\
\citet{theo46} & total,$nl$ & $\rm B^{2+} + H $ & $\sim$ 10$^{-1}$ - 10$^4$  \\
\citet{theo35} & total,$nl$ & $\rm B^{4+} + H$  & 10$^{-2}$ - 2$\times 10^5$ \\
\citet{theo2} & total & $\rm C^{+} + H$ & $\leq$ 10$^{3}$ \\
\citet{theo5} & total & $\rm C$, $\rm N$, $\rm O$, $\rm Si + H^{+}$ & $\leq$ 10$^3$ \\
\citet{theo15} & total & $\rm C^{q+} $, $\rm N^{q+} $, $\rm O^{q+} $, $\rm Ne^{q+} + H$ (q = 2,3), & 10$^{-1}$ - 1  \\
 &  & $\rm O^{2+} + He$  & \\
 \citet{theo16} & total & $\rm C^{q+} $, $\rm N^{q+} $, $\rm O^{q+} $, $\rm Ne^{q+}$, $\rm Mg^{q+}$,  $\rm Si^{q+}$, $\rm S^{q+}$,  $\rm Ar^{q+} $  & 10$^{-2}$ - 1 \\
 &  & $+ H$ (q = 2,3,4)  &  \\
 \citet{theo19} & $nl$ & $\rm C^{2+}, N^{3+}, O^{3+}, Ne^{2+}, Ne^{3+} + H$ & 0.27 - 8.1 \\
 \citet{theo33} & total & $\rm C^{3+}, O^{3+}, Si^{3+} + H$  & $\sim$ 0.1 - 10$^4$ \\
 \citet{theo41} & total,$nl$ & $\rm C^{3+} + He $ & $\sim$ 10$^{-4}$ - 10$^3$ \\
 \citet{theo65} & total  & $\rm C^{q+}, N^{q+}, O^{q+}, F^{q+}, Ne^{q+}, Kr^{q+} $ & 600  \\
& & + H (q = $10-25$ for Kr and $4-9$ for rest) & \\
\citet{obs37} & total,$nl$ & $\rm C^{6+}, N^{7+}, O^{8+}, Ne^{10+}, Na^{11+}, Mg^{12+},$  & $\sim$ 10$^{-3}$ - 10$^{5}$ \\
 &  & $\rm Al^{13+}  \rm + H, H_{2}O, CO, CO_{2}, OH, O$  &  \\
\citet{theo73} & total,$nl$ & $\rm C^{6+} + H, He $ & 2.7$\times 10^{3}$ - 8.3$\times 10^{3}$ \\
\citet{theo74} & total,$nl$ & $\rm C^{6+} - Al^{13+} + H, He, H_{2} $ & 2 $\times 10^2$ - 5 $\times 10^3$ \\
\citet{theo24} & total,$nl$ & $\rm N + H^{+}, N^{+} + H$  & $\sim$ 10$^{-4}$ - 10$^3$  \\
\citet{theo28} & total & $\rm N^{2+} + H$  & 2$\times 10^{-3}$ - 3 $\times 10^{5}$ \\
\citet{theo61} & total & $\rm N^{2+} + H $ & $\sim$ 10$^{-1}$ - 10$^2$ \\
\citet{theo54} & total,$nl$ & $\rm N^{4+} + H $ & $\sim$ 50 - 2$\times 10^4$ \\
\citet{theo64} & total  & $\rm N^{4+} + H $ & 30 - 10$^5$ \\
\citet{theo58} & total,$nl$ & $\rm N^{4+} + H $ & $\sim$ 0.1 - 8$\times 10^3$ \\
\citet{theo59} & total,$nl$ & $\rm N^{4+} + H $ & $\sim$ 10$^{-2}$ - 6$\times 10^3$ \\
\citet{theo25} & total,$nl$ & $\rm N^{5+} + H$  & $\sim$ 10$^{-2}$ - 10$^4$ \\
\citet{hasan2001} & $n$ & $\rm N^{7+}, O^{7+} + He, CO, CO_{2}, H_{2}O$  & 2$\times 10^3$ - 4.67$\times 10^3$ \\
\citet{theo6} & total & $\rm O + H^{+}$ & $\leq 10^{-1}$ \\
\citet{theo11} & total & $\rm O + H^{+}$, $\rm O^{+} + H$ & 10$^{-4}$- 10$^7$ \\
\citet{theo69} & total,$nl$ & $\rm O^{+} + H_{2}$ & 1 - 10$^4$ \\
\citet{theo71} & total,$nl$ & $\rm O^{+} + H_{2} $ & $\sim$ 5$\times 10^2$ - 10$^4$  \\
\citet{theo27} & total & $\rm O^{2+} + H, N^{2+} + H$  & $\sim$ 10$^{-3}$ - 10$^44$  \\
\citet{theo29} & total & $\rm O^{2+} + H$  & $\sim$ 125 - 3.4 $\times 10^3$ \\
\citet{theo31} & total,$nl$ & $\rm O^{2+} + H$  & $\sim$ 10 - 3$\times 10^4$  \\
\citet{theo60} & total & $\rm O^{2+} + H $ & $\sim$ 3$\times 10^{-4}$ - 4  \\
\citet{theo13} & total,$nl$ & $\rm O^{3+} + H$ & 0.1 - 10$^3$  \\
\citet{theo14} & total & $\rm O^{3+} + H$ & 0.1 -5$\times 10^3$ \\
\citet{theo17} & $nl$ & $\rm O^{3+} + H$ & 0.1 - 1 \\
\citet{theo18} & $nl$ & $\rm O^{3+} + H$ & $\leq 1$ \\
\citet{theo21} & total & $\rm O^{3+} + H$  & $\sim$ 1  \\
\citet{theo42} & total,$nl$ & $\rm O^{3+} + He $ & 0.01 - 10$^3$  \\
\citet{theo70} & total,$nl$ & $\rm O^{3+} + H_{2} $ & 0.1 - 10$^4$ \\
\citet{theo72} & total,$nl$ & $\rm O^{6+} + H$ &  0.1 - 10$^4$  \\
\citet{theo43} & total,$nl$ & $\rm Ne^{2+} + He $ & 0.1 - 10$^4$  \\
\citet{theo38} & total,$nl$ & $\rm Ne^{(8-10)+} + H , Mg^{(8-12)+} + H$  & 10$^{-3}$ - 5 $\times 10^4$  \\
\citet{theo68} & $nl$,line & $\rm Ne^{10+} + Ne, He, H_{2}, CO_{2}, H_{2}O$  & 9  \\
\citet{theo7} & total & $\rm Na + H^{+}$ & $\leq$34  \\
\citet{theo8} & total & $\rm Mg + H^{+}$ & $\leq$40  \\
\citet{theo37} & total,$nl$ & $\rm Mg^{12+} + H, He$ & 10$^3$ - 5 $\times 10^3$  \\
\citet{theo9} & total & $\rm Si + H^{+}$ & $\leq$30 \\
\citet{theo52} & total & $\rm Si^{2+} + He $ & $\sim$ 20 - 6 $\times 10^3$ \\
\citet{theo57} & total,$nl$ & $\rm Si^{2+} + H $ & $\sim$ 0.03 - 100 \\
\citet{theo44} & total,$nl$ & $\rm Si^{3+} + H $ & 0.01 - 10$^{6}$ \\
\citet{theo53} & total,$nl$ & $\rm Si^{3+} + He $ & $\sim$ 10$^{-4}$ - 400 \\
\citet{theo55} & total,$nl$ & $\rm Si^{3+} + He $ &  1 - 7 $\times 10^{3}$ \\
\citet{theo50} & total & $\rm Si^{3+} + He $ & $\sim$ 10$^{-4}$ - 10$^3$  \\
\citet{theo34} & total,$nl$ & $\rm Si^{3+} + H$  & 0.01 - $2\times 10^{5}$ \\
\citet{theo51} & total & $\rm Si^{4+} + He $ & 10$^{-3}$ - 2.5 $\times 10^{3}$  \\
\citet{theo47} & total,$nl$ & $\rm Si^{4+} + He $ & $\sim$ 10$^{-4}$ - 10$^7$ \\
\citet{theo49} & total,$nl$ & $\rm Si^{4+} + He $ & $\sim$ 0.01 - 100  \\
\citet{theo32} & total,$nl$ & $\rm P^{2+} + H$  & $\sim$ 100 - 10$^{5}$  \\
\citet{theo10} & total & $\rm S + H^{+}$ & $\leq$ 10$^3$ \\
\citet{theo23} & total,$nl$ & $\rm S + H^{+}$  & $\sim$ 10$^{-4}$ - 10$^4$  \\
\citet{theo63} & total,$nl$ & $\rm S^{2+} + H $ &  100 - 10$^{4}$  \\
\citet{theo45} & total,$nl$ & $\rm S^{3+} + H $ &  2$\times 10^3$ - 8$\times 10^3$  \\
\citet{theo56} & total,$nl$ & $\rm S^{3+} + He $ &  2$\times 10^3$ - 5$\times 10^4$ \\
\citet{theo48} & total,$nl$ & $\rm S^{4+} + H $ & $\sim$ 10$^{-4}$ - 10$^7$  \\
\citet{theo40} & total,$nl$ & $\rm Fe^{25+}, Fe^{26+} + H $ & 10$^{-3}$ - 10$^5$ \\
\citet{theo36} & total,$nl$ & $\rm Ge^{q+}, Se^{q+}, Br^{q+}, Kr^{q+}, Rb^{q+}, Xe^{q+}$   & 10$^{-3}$ - 10$^6$ \\
 &  & $\rm + H$ (q = 2,3,4,5)   &  \\

\hline

\end{longtable}
\noindent a: total = total cross section, nl = $nl$-resolved cross section \\

To produce accurate X-ray emission models, it is desirable to have absolute cross sections using state-resolved theoretical calculation for all the relevant reactions at a broad range of collision velocities. Simple approximations described above, e.g., COB, cannot reproduce accurate experiment results (e.g., \citealt{hasan2001}). Furthermore, COB-like models do not have the facility to calculate the $l$ distribution.

For collisions with energies $E \geq 5$ keV/u, the Classical Trajectory Monte Carlo (CTMC) method is often considered to be a valid approximation (e.g., \citealt{exp59}). The CTMC provides a non-perturbative solution to the classical equations of motion in the many body interactions during the collisions. However, this method does not give an adequate description of the lowering of $l$ quantum number as the collision energy decreases below $\sim 1$~keV \citep{peter2000}. 

The Multichannel Landau-Zener approach (MCLZ) provides another flexible tool for massive calculation. MCLZ considers a slow collision between an ion and a neutral particle using a quasi-molecular configuration, therefore, it is mostly used for models with $E \leq$ a few keV/u. It studies the crossing between pre- and post-exchange potentials of the configuration and determines the final state of the capture. Since the MCLZ requires that crossing distances are fully resolved, while for a bare ion collision, the capture $l$ degenerates with the $n$ state, this method cannot resolve $l$ distribution for bare ion collision \citep{theo40}. Empirical $l$ distribution functions listed in \ref{sec:physics} must then be applied.

(Quasi-)quantum mechanical models involving close coupling are known to be accurate simulations of atomic collisions. The Atomic-Orbital Close-Coupling (AOCC) is found to be applicable between $\sim 0.1$ and $100$ keV/u. It is based on bound state calculation on either center of the two reactors, assuming an infinite separation between them. The number of $n$ and $l$ are limited by the finite size of the close-coupling expansion.

The Molecular-Orbital Close-Coupling (MOCC) is another type of close-coupling which is optimized for low energy ($E \leq 10$ keV/u). It models the union state of the two reactors. The molecular wave functions, as well as the coupling terms of the relevant status, are calculated and fed into the close-coupling scatter calculation. The number of $n$ and $l$ are limited in the same way as the atomic-orbital. A full quantum mechanical version, QMOCC, has been developed to cope with very low energy collision at $E < 10$ eV/u \citep{stancil01}.

We summarize in Table~\ref{theo_data} a comprehensive, though not exhaustive, list of theoretical calculations. The reactors, type of the data products (total, $n-$, and $nl-$ resolved cross sections) and the energy range indicated in the references, are all recorded. The collision energy determines the range of application; for instance, the solar wind ions collide with comet atmosphere at $E \sim 0.1 - 5$ keV/u, while for the charge exchange component in collisional or photo-ionization equilibrium, the relevant collisions are much less energetic with $E \sim 10$ eV/u.

\subsection{Laboratory measurements}

\onecolumn
\begin{longtable}[H]{c | c | c | c }
\caption{Experimental cross section data} \\
\label{exp_data}\\
\centering
Reference &  type$^{a}$  & ion & energy (eV/u)  \\
\endfirsthead
\centering
\centering
Reference &  type$^{a}$  & ion & energy (eV/u)  \\
\hline
\endhead
\endfoot
\hline
\citet{exp39} & total & $\rm H^{+}, He^{+}, He^{2+} + H_{2}O, CO_{2} $ &  300 - 7.5$\times$10$^{3}$ \\
\citet{exp42} & nl & $\rm He^{2+}, N^{5+} + H_{2}$  & 5 - 10$^{3}$ \\
\citet{exp57} & total & $\rm He^{2+} +  H_{2}, N_{2}, O_{2}, CO, CO_{2}$  &  4$\times$10$^{3}$ \\
\citet{exp68} & total & $\rm He^{2+} + He$ &  2$\times$10$^{2}$ - 4$\times$10$^{3}$ \\
\citet{exp1} & total & $\rm Li^{q+}$ (q=1-3) & 4$\times$10$^{3}$-3.43$\times$10$^{5}$ \\
\citet{exp2} & total & $\rm Li^{q+}$ (q=2-3), $\rm N^{q+}$ (q=2-5), $\rm Ne^{q+}$ (q=3-5) &  3q$\times$10$^{3}$ - 1.2q$\times$10$^{4}$ \\
\citet{1979JPhB...12.3763G} & total & $\rm B^{q+}$ (q=1-5), $\rm C^{q+}$ (q=1-4) &  q$\times$10$^{5}$ - 2.5q$\times$10$^{6}$ \\
 & total & $\rm C^{q+}$ (q=5,6), $\rm N^{7+}$ & q$\times$10$^{5}$ - 2.5q$\times$10$^{6}$ \\
\citet{exp3} & total & $\rm B^{2+}$, $\rm C^{+}$, $\rm N^{+}$, $\rm Mg^{2+}$ & 8$\times$10$^{2}$ - 4$\times$10$^{4}$ \\
\citet{exp4} & total & $\rm B^{q+}$ (q=2-5), $\rm C^{q+}$ (q=3,4), $\rm N^{q+}$ (q=3,4), $\rm O^{q+}$ (q=5,6) &  4q$\times$10$^{3}$ - 2.5q$\times$10$^{4}$\\
\citet{exp25} & total & $\rm B^{4+}$ & 60 - 1.2$\times$10$^{3}$ \\
\citet{exp5} & total & $\rm B^{q+}$ (q=2-4), $\rm C^{q+}$ (q=2-4), $\rm N^{q+}$ (q=2-5), $\rm O^{q+}$ (q=2-5) & 6q$\times$10$^{3}$ - 2.3q$\times$10$^{4}$ \\
\citet{1979PhRvA..19..515M} & total & $\rm B^{q+}$ (q=2-5), $\rm C^{q+}$ (q=3,4), $\rm N^{q+}$ (q=3,4) & 2.5$\times$10$^{4}$ - 2$\times$10$^{5}$ \\
\citet{exp6} & total & $\rm C^{q+}$ (q=1-4), $\rm N^{q+}$ (q=1-5), $\rm O^{q+}$ (q=1-5), $\rm Si^{q+}$ (q=2-7) & 8.6$\times$10$^{3}$ - 1.65$\times$10$^{6}$ \\
\citet{exp7} & total & $\rm C^{2+}$ & 5$\times$10$^{2}$ - 1.4$\times$10$^{3}$ \\
\citet{exp23} & total & $\rm C^{3+}$ & 0.3 - 3$\times$10$^{3}$ \\
\citet{exp44} & total,line & $\rm C, N, O, Ne^{q+} + He, H_{2}, H_{2}O, CO_{2}$ (q = 3-9)  &  7q$\times$10$^{3}$  \\
\citet{exp40} & total,line & $\rm C, N, O, Ne^{q+} + H_{2}O, CO_{2}$ (q = 3-10)  &  550 - 800  km s$^{-1}$ \\
\citet{exp72} & line & $\rm C^{2+} + H_{2}$ &  10$^{3}$ - 3$\times$10$^{4}$ \\
\citet{1982PhRvA..26.1892P} & total & $\rm C^{q+}$ (q=3,4), $\rm O^{q+}$ (q=2-6) & 10 - 1$\times$10$^{4}$ \\
\citet{exp41} & nl & $\rm C^{4+}, N^{5+}, O^{6+} + H_{2}$  & 5 - 4$\times$10$^{3}$  \\
\citet{1982PhRvA..26.1892P} & total & $\rm C^{q+}$ (q=5,6)& 10 - 1$\times$10$^{4}$ \\
\citet{exp8} & total & $\rm C^{3+}$ & 1$\times$10$^{6}$ - 3.5$\times$10$^{6}$ \\
\citet{exp9} & total,nl & $\rm C^{q+}$ (q=3,4), $\rm N^{5+}$, $\rm O^{6+}$ & 7$\times$10$^{2}$ - 4.6$\times$10$^{3}$ \\
\citet{exp10} & total,nl & $\rm C^{3+}$ & 6$\times$10$^{2}$ - 1.8$\times$10$^{4}$ \\
\citet{1983PhST....3..124P} & total & $\rm C^{4+}$, $\rm N^{5+}$, $\rm O^{6+}$, $\rm Ne^{8+}$ & 100 - 1300 km s$^{-1}$ \\
 & total & $\rm C^{q+}$ (q=5,6), $\rm N^{q+}$ (q=6,7), $\rm O^{q+}$ (q=7,8), $\rm Ne^{q+}$ (q=9,10) & 100 - 1300 km s$^{-1}$  \\
\citet{exp11} & total,nl & $\rm C^{q+}$ (q=3,4), $\rm N^{5+}$, $\rm O^{6+}$ & 2.7 - 13.6  \\
\citet{exp52} & total & $\rm C^{q+} + CO$ (q = 3-4)  &  5$\times$10$^{2}$ \\
\citet{exp12} & total,nl & $\rm C^{4+}$ & 100 - 2$\times$10$^{4}$ \\
\citet{exp13} & total,nl & $\rm C^{4+}$ & 2.4$\times$10$^{2}$ - 4.4$\times$10$^{2}$  \\
\citet{exp58} & total & $\rm C^{4+} + He$  &  2.4$\times$10$^{2}$ - 4.4$\times$10$^{2}$ \\
\citet{exp66} & total & $\rm C^{5+} + H$ &  0.64 - 1.2$\times$10$^{4}$ \\
\citet{exp83} & total, n & $\rm C^{5+} + H_{2}O$  & 113 - 3.75$\times$10$^{3}$ \\
\citet{exp51} & line & $\rm C^{q+}, O^{q+} + CO_{2}$ (q = 5-8)  &  several 100 km s$^{-1}$ \\
\citet{exp70} & line & $\rm C^{6+} + He$ &  460 - 3.2$\times$10$^{4}$  \\
\citet{exp77} & total,nl & $\rm C^{6+} + He, H_{2}$  & 10$^{3}$ - 2.5$\times$10$^{4}$ \\
\citet{exp74} & line & $\rm C^{6+} + Kr$ &  320 - 4.6$\times$10$^{4}$  \\
\citet{exp80} & line & $\rm C^{6+}, O^{8+} + Kr$  & 500 - 4$\times$10$^{4}$  \\
\citet{exp14} & total & $\rm N^{+}$, $\rm O^{+}$ & 100 - 4$\times$10$^{4}$  \\
\citet{exp26} & total & $\rm N^{4+}$ & 1-3$\times$10$^{2}$  \\
\citet{exp34} & total & $\rm N^{4+}$ + H , $\rm H_{2}$ &  10$^{3}$-4$\times$10$^{3}$  \\
\citet{exp36} & total & $\rm N^{q+}$ + H (q = 3-5)&  0.9 - 1.4$\times$10$^{3}$ \\
\citet{exp86} & total & $\rm N^{7+}, O^{7+} + H$  & 2 - 1.22$\times$10$^{3}$ \\
\citet{hasan2001} & total,n & $\rm N^{7+}, O^{7+} + He, H_{2}O, CO_{2}, CO$  & 2$\times$10$^{3}$, 4.67$\times$10$^{3}$ \\
\citet{exp15} & total & $\rm O^{+}$ & 100 - 3.6$\times$10$^{4}$\\
\citet{exp47} & total & $\rm O^{+} + H_{2}, CO_{2}, CO, CH_{4}, C_{2}H_{2}, C_{2}H_{6}, C_{3}H_{8}$  &  200 - 4.5$\times$10$^{3}$ \\
\citet{exp19} & nl & $\rm O^{3+}$ & 45 - 752  \\
\citet{exp28} & total & $\rm O^{q+}$ (q = 2,3) + H & 1.5$\times$10$^{4}$ \\
\citet{exp29} & total & $\rm O^{q+}$ (q = 3,4) + H(D) & 1-10$^{3}$ \\
\citet{1979PhRvA..19..515M} & total & $\rm O^{q+}$ (q=3-6), $\rm Si^{q+}$ (q=4-9), $\rm Fe^{q+}$ (q=4-15) & 2.5$\times$10$^{4}$ - 2$\times$10$^{5}$ \\
\citet{exp48} & total & $\rm O^{q+} + CO$ (q = 4,5)  &  1.5q$\times$10$^{3}$ \\
\citet{exp16} & total & $\rm O^{5+}$ & 0.9-800  \\
\citet{exp62} & total,nl & $\rm O^{6+} + H_{2}O$  &  100 -7.5$\times$10$^{3}$ \\
\citet{exp67} & line & $\rm O^{6+} + CO$ &  3.6$\times$10$^{4}$ \\
\citet{exp76} & total & $\rm O^{6+} + CO, H_{2}O, CO_{2}, CH_{4}, N_{2}, NO, N_{2}O, Ar$  &  1.17$\times$10$^{3}$ - 2.33$\times$10$^{3}$ \\
\citet{1979PhRvA..19..515M} & total & $\rm O^{q+}$ (q=7,8) & 2.5$\times$10$^{4}$ - 2$\times$10$^{5}$ \\
\citet{exp43} & line & $\rm O^{8+}, O^{7+} + CO_{2} $  & 15  \\
              &      & $\rm Ne^{10+}, Ne^{9+} + Ne $   & 9  \\
\citet{exp37} & total & $\rm O^{q+}$ + He (q = 2-5)&  100 -1.5$\times$10$^{3}$ \\
\citet{exp84} & total, n & $\rm O^{6+} + CO_{2}, CH_{4}, H_{2}, N_{2}$  & 2.63$\times$10$^{3}$ - 1.95$\times$10$^{4}$ \\
\citet{exp59} & line & $\rm O^{8+} + CH_{4}, CO_{2}, N_{2}$  &  10  \\
\citet{exp79} & line & $\rm O^{8+} + Kr$  & 445 - 8.18$\times$10$^{3}$ \\
\citet{exp69} & total, n & $\rm O^{8+} + He$ &  16 - 54  \\
\citet{exp24} & total & $\rm Ne^{2+}$ & 139 - 1.49$\times$10$^{3}$ \\
\citet{exp30} & total & $\rm Ne^{2+}$ + H(D) & 59-949 \\
\citet{exp17} & total & $\rm Ne^{q+}$ (q=2-4), $\rm Ar^{q+}$ (q=2-4,6) & 10$^{3}$ - 1.5$\times$10$^{4}$ \\
\citet{exp32} & total & $\rm Ne^{3+}$ + H & 0.07 -826 \\
\citet{exp75} & total,n & $\rm Ne^{q+} + CO_{2}, H_{2}O$ (q = 3-5) &  15q - 500q \\
\citet{exp31} & total & $\rm Ne^{4+}$ + H & 0.1 - 1006  \\
\citet{exp71} & total& $\rm Ne^{q+} + Ne, Ar, Kr, Xe$ &  2q$\times$10$^{3}$  \\
\citet{exp65} & total & $\rm Ne^{6+} + CO_{2}, H_{2}O$ &  450 - 2.4$\times$10$^{3}$ \\
\citet{exp88} & line & $\rm Ne^{8+} + He, Kr$  & 400-900 km s$^{-1}$ \\
\citet{exp82} & total, n & $\rm Ne^{8+} + He, H_{2}$  & 4.6 - 10.9  \\
\citet{exp85} & total, n & $\rm Ne^{8+}, Ne^{9+} + He, H_{2}$  & 10$^{3}$ - 2.475$\times$10$^{4}$ \\
\citet{exp73} & line & $\rm Ne^{10+}, Ar^{18+}, Kr^{36+} + Ar$ &  18 - 4$\times$10$^{3}$ \\
\citet{exp55} & line & $\rm Kr^{36+}, Ar^{18+}, Ne^{10+} + Ar$  &  4$\times$10$^{3}$ \\
\citet{exp18} & total & $\rm Si^{q+}$ (q=2-7) & 5.1$\times$10$^{4}$ - 2.04$\times$10$^{6}$  \\
\citet{exp22} & total & $\rm Si^{3+}$ & 40 - 2500 \\
\citet{exp33} & total & $\rm Si^{q+}$ + He (q = 3-5) & $\sim$ 100 - 1000 \\
\citet{exp45} & total,line & $\rm Si^{13+}, S^{15+}, Ar^{17+} + Ar, He, CH_{4}$  & 10$^{3}$-7$\times$10$^{4}$ \\
\citet{exp35} & total & $\rm S^{2+}$ + H , $\rm H_{2}$ &  2$\times$10$^{3}$-8$\times$10$^{3}$ \\
\citet{exp78} & line & $\rm S^{16+} + CS_{2}$  & $\sim 10$ \\
\citet{exp38} & total,nl & $\rm Ar^{q+}$ + H, $\rm H_{2}$, He (q = 4-6)&  q$\times$10$^{3}$, 2q$\times$10$^{3}$ \\
\citet{exp87} & total, nl & $\rm Ar^{8+} + He$  & 1.4$\times$10$^{3}$ - 2$\times$10$^{4}$ \\
\citet{exp54} & line & $\rm Ar^{16+} + H$  &  4$\times$10$^{4}$  \\
\citet{exp60} & line & $\rm Ar^{17+}, Ar^{18+} + Ar$  &  0.25q - 125q \\
\citet{exp64} & line & $\rm Ar^{17+}, Ar^{18+} + Ar$ &  5 - 2$\times$10$^{3}$ \\
\citet{exp63} & line & $\rm Fe^{q+} + O_{2}, N_{2}, H_{2}O, CO_{2}$ (q = 16 - 23) &  $\leq$ 10-20 \\
\citet{exp56} & total & $\rm Cu^{q+} +  H_{2}, Ne, Ar, Kr, Xe$  &  2.5$\times$10$^{6}$\\
\citet{exp81} & line & $\rm Ni^{19+}, Ni^{18+} + He, H_{2}$  & 10  \\
\citet{exp49} & total,line & $\rm Kr^{q+} + He, Ne, Ar, N_{2}, CH_{4}$ (q = 27-33)  &  2$\times$10$^{3}$ \\
\citet{exp61} & line & $\rm Xe^{18+}, Xe^{24+} + Na$  &  500 - 4.5$\times$10$^{3}$ \\
\citet{exp50} & total,line & $\rm Ta^{q+} + Ar$ (q = 45-49)  &  10$^{3}$ - 4$\times$10$^{4}$ \\

\hline
\end{longtable}
\noindent a: total = total cross section, n = $n$-resolved cross sections, nl = $nl$-resolved cross sections, line = line emissivities or ratios \\

\begin{figure}[bht]
\sidecaption
\includegraphics[width=\textwidth]{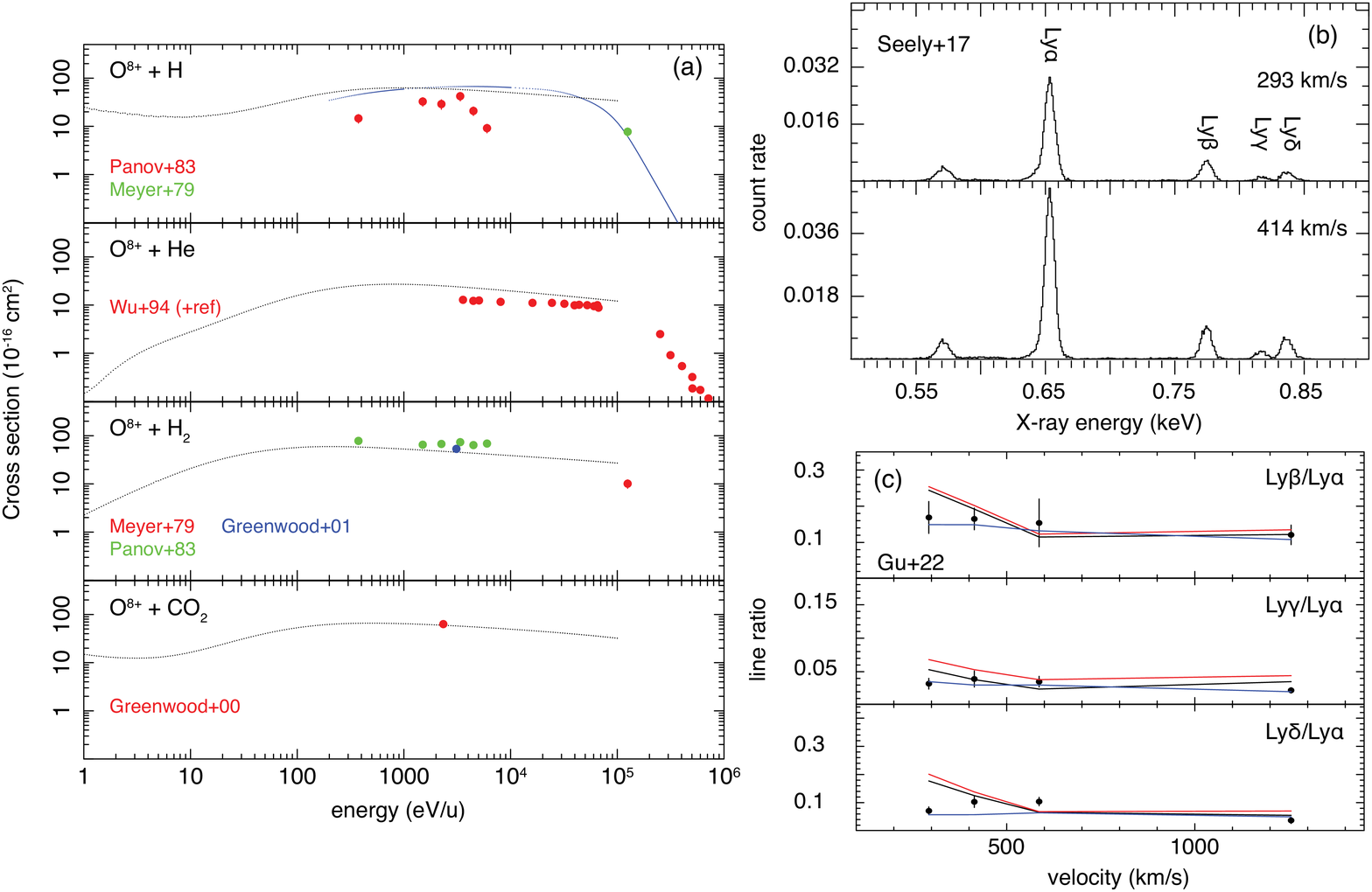}
\caption{(a) Total cross sections as a function of collision energy for O$^{8+}$ interacting with atomic hydrogen, atomic helium, molecular hydrogen, and CO$_{2}$. The solid lines are the model values from the calculations with the MCLZ (black) and \citet{janev1993}. Data points are experimental results from sources indicated in the labels. (b) X-ray spectra from \citet{exp79} on O$^{8+}$ + Kr reaction for two different velocities. (c) Comparison of experimental and theoretical line ratios for O$^{8+}$ charge exchange. The experimental data are taken from \citet{exp79}, and the predictions from \citet{gu2016a}, MCLZ, and \citet{janev1993} are shown in black, red, and blue curves. }
\label{fig:o8}       
\end{figure}

Laboratory measurements provide the absolute calibration of the atomic constants. The charge exchange process has been measured in the laboratory since 1950s (e.g., \citealt{hasted1951}), using a crossed-beam setup in which a beam of ionized particles passes through a chamber filled with neutral gas. The storage ring facilities are often used for measurements at a high collision energy ($E \sim$ MeV/u), while the electron cyclotron resonances and the electron beam ion sources are applied to measurements with medium collision energy ($E \sim$ keV/u). A significant part of the early studies with crossed-beam measurements contributed to the fusion research because of the charge exchange influence on plasma parameters such as charge state distribution and radiative cooling {e.g., \citealt{exp2}}. Recent interest arises for astrophysical applications with the measurements of highly charged ion collisions, including among many others, the total cross sections of bare C, N, O, and Ne charge exchanging with H$_2$O, He, H$_2$, CO$_2$ measured at Jet Propulsion Laboratory \citep{exp39, exp40}, and the K-shell charge exchange between He$^{2+}$ and H$_2$O using the crossed-beam facility at Kernfysisch Versneller Instituut in Groningen \citep{exp13,bodewits2005}. 

The combination of charge state determination and spectroscopic measurements becomes more available lately, providing reliable measurement on the velocity-dependent total and state-resolved cross sections. Some such measurements has been made with the cold target recoil-ion momentum spectroscopy (COLTRIMS), which has been developing since 1990s (e.g., \citealt{abda1998}). The $n$ states, sometimes even $l$ levels of the captured electrons, are inferred from the measurement of the ion kinetic energy loss (the ``Q'' value, e.g., \citealt{fischer2002,knoop2008,xue2014,zhang2017}). A few most recent efforts are carried out in the COLTRIMS facilities in the Institute of Modern Physics of China on the Ne$^{8+}$ and Ne$^{9+}$ collision with He and H$_2$ \citep{exp85}, and at the COLTRIMS in combination with simultaneous X-ray spectroscopic experiments at the electron cyclotron resonance source of the University of Nevada, Reno \citep{ali2016}.

One potential issue with beam-gas approach is that the radiative cascade from the metastable levels of the product ions might take longer than the dynamical timescale in the interaction region, making it complicated to study the X-ray spectra observed. The electron beam ion trap (EBIT)~\citep{levine1988} devices provide a solution, as the ions are radially confined in a quasi-stationary state by the space charge potential of the electron beam and axially trapped by the electrostatic potential to the cylindrical drift tube electrodes. A useful technique for measuring charge exchange with the EBIT is the magnetic trap mode, in which the electron beam is turned off when the proper charge states are reached. The ions expand to form a larger cloud, but remain confined in the radial direction by the strong magnetic field and the potential applied to the trap electrodes. Typically, the trap potentials limit the energy of the ions stored in the EBIT. Therefore, the collision energy for the CX reaction is usually very low, less than 10--20 eV/u~\citep{peter2000}. However, the advantage of the magnetic trapping mode is that the electron-ion collision can be safely eliminated, while the ion-neutral collision remains efficient and produces the pure charge exchange emission.

The experiments using EBITs do not measure the absolute cross sections; instead, they often record the detailed X-ray spectra of the reactions. Earlier EBIT experiments with limited spectral resolution used the parameter known as the hardness ratio, i.e., the intensity ratio of X-ray line emission from levels with $n\geq3$ relative to $n=2$ K-shell emission, to study the dependence on collision energies~\citep{peter2000,exp64,exp63,exp78}. Recent high-resolution (resolved to a few eV) X-ray spectral measurements have become possible with the introduction of X-ray microcalorimeters~\citep{bei2003}. The high resolution is particularly important for charge exchange, which emits lines from high $n$ levels, often in a narrow energy range. The NASA Goddard Space Flight Center X-ray Spectrometer (XRS) and EBIT Calorimeter Spectrometer (ECS)~\citep{porter2008xrs,porter2008ecs} have been used for a number of charge exchange measurements performed at the Lawrence Livermore National Laboratory EBIT (e.g., \citealt{frankel2009,leutenegger2010}).

The $n$-resolved line diagnostics are made possible with the EBIT plus high-resolution spectrometer, while the $l$-distribution of the capture, which is sensitive to the collision velocity (see \S~\ref{sec:physics}), remains challenging since the ion-neutral collisions in the magnetic trapping mode are always mild ($\leq 10$ eV). A possible solution is to measure the charge exchange photons simultaneously at longer wavelengths, e.g., in the UV and optical bands, where the radiative cascades of the Rydberg levels might provide additional constraints on the $l$-state distribution. In addition, efforts are needed to combine EBIT and high-resolution calorimeters with the COLTRIMS setup to fully investigate the kinematics of charge-exchange collision and multi electron capture processes~\citep{ali2016}. Furthermore, the development of a suitable atomic hydrogen target for EBIT measurements is ideal so that only single-electron capture capture can take place~\citep{leutenegger2016}.

In Figure~\ref{fig:o8}, we show a set of historical and state-of-the-art measurements of the O$^{8+}$ charge exchange. The total cross sections of O$^{8+}$ with H, He, H$_2$, and CO$_2$ targets are measured in crossed-beam in the 1970s to 2000s. In panel b, the X-ray quantum microcalorimeter spectra taken with the beam-gas measurements at Oak Ridge on O$^{8+}$ with Kr charge exchange are shown for two different velocities. The comparison of the observed and calculated line ratios is plotted in panel c. A more comprehensive list of laboratory measurement efforts made in the past can be found in Table~\ref{exp_data}.

\section{Observations}
\label{sec:3}
\subsection{Ionization balance}

\begin{figure}[bht]
\sidecaption
\includegraphics[width=\textwidth]{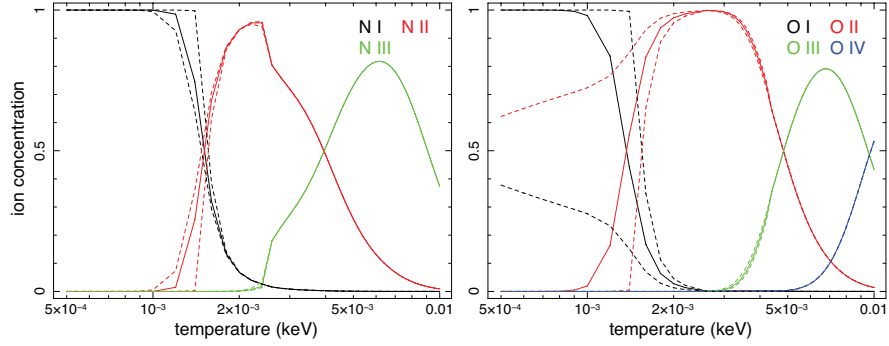}
\caption{Charge state distributions of N (left) and O (right) as a function of equilibrium temperature for the CIE plasma. The dashed lines show the calculations
when the charge exchange recombination rates are changed by 50\%, while the other ionization and
recombination data are kept the same.}
\label{fig:cxcon}       
\end{figure}

Charge exchange is relevant for determining the ion concentration as it serves as a recombination term for ion $X^{q+}$ with 
\begin{equation}
    X^{q+} + H (He) \rightarrow X^{(q-1)+} + H^{+} (He^{+}), 
\end{equation}
but also in some conditions an ionization term with
\begin{equation}
    X^{(q-1)+} + H^{+} (He^{+}) \rightarrow X^{q+} + H (He). 
\end{equation}
These terms are most effective for photoionized and non-equilibrium collisional ionization plasmas where highly charged ions naturally co-exist with the neutral hydrogen and helium. As shown in \citet{gu2022}, the charge exchange contribution could even be effective for the collisional ionization equilibrium at a low temperature ($\sim$ a few eV, Fig.~\ref{fig:cxcon}). The collision induced by the thermal motion of ions has a typical velocity of km s$^{-1}$. 

Sets of recombination and ionization rates by charge exchange are tabulated in \citet{ar1985} and \citet{kf1996}. These tables have been used broadly in the ion concentration calculation, however, they lacks data of highly charged ions with $q > 4$. Moreover, a significant part of the rates are derived from theoretical calculations with classical approximations, e.g., Landau-Zener method. A systematic review and update are then needed to expand the $q$ range as well as to improve the quality of charge exchange data for the ionization balance calculation.

In table~\ref{obs} we list a set of astrophysical observations made so far. For each observation, we feature the object, instrument, relevant emission features, together with a few remarks on the quality of the detection (e.g, using image only, with timing, and using spectrum with normal or high spectral resolution).  

\onecolumn
\begin{longtable}[H]{c | c | c | c | c}
\caption[longtable]{Observations of charge exchange in X-ray}\\
\label{obs}\\
\centering
Reference & detector & sources & relevant ion(s) & note  \\
\endfirsthead
\centering
\centering
Reference & detector & source(s) & relevant ion(s) & note  \\
\hline
\endhead
\endfoot
\hline
\citet{obs1} & ROSAT   & C/Hyakutake 1996 B2    & $-$ & image\\
\citet{obs2} & BeppoSAX & C/1995 O1 (Hale-Bopp) & $-$ & image, spectrum \\
\citet{obs3} & ROSAT, EUVE   & P/Encke 1997    & $-$ & image\\
\citet{obs4} & Chandra   & Mars   & $-$ & spectrum, low S/N \\
\citet{obs5} & Chandra   & C/1999 T1  & $-$ & image, spectrum\\
\citet{obs6} & Chandra   & Earth geocorona   & O VII, O VIII & spectrum\\
\citet{obs7} & XMM-Newton   & Heliosphere  & C, O, Ne, Mg & spectrum\\
\citet{obs8} & Chandra   & 2P/Encke 2003  & $-$  & image, spectrum\\
\citet{obs9} & XMM-Newton   & Mars  & C, N, O, Ne  & high res. spectrum\\
\citet{obs10} & Swift & 9P/Tempel 1 (deep impact) & $-$ & timing, spectrum \\
\citet{obs11} & Chandra & sample of 4 comets  & $-$ & spectrum \\
\citet{obs12} & Suzaku & Earth geocorona  & $-$ & spectrum \\
\citet{obs13} & Chandra & Venus  & $-$ & spectrum, low S/N \\
\citet{obs14} & Chandra & sample of 8 comets  & $-$ & spectrum \\
\citet{obs15} & XMM-Newton & Earth geocorona & O VII & timing, spectrum \\
\citet{obs16} & Chandra & 73P/2006, C/1999 S4 & $-$ & spectrum \\
\citet{obs17} & Chandra & 8p/Tuttle, 17p/Holmes & $-$ & spectrum \\
\citet{obs18} & Suzaku & Earth geocorona & O VII variation & timing, spectrum \\
\citet{obs19} & XMM-Newton & Heliosphere & O VII, O VIII & spectrum \\
\citet{obs20} & Suzaku & Cygnus Loop SNR & O VII at 0.7~keV & spectrum \\
\citet{obs100}& XMM-Newton & Puppis A    & O VII         & high res. spectrum \\  
\citet{obs21} & XMM-Newton & M82 & O VII, Ne IX, Mg XI & high res. spectrum \\
\citet{obs22} & Suzaku & M82 & O VIII, Ne X & spectrum \\
\citet{obs23} & XMM-Newton & 9 star-forming galaxies & O VII & high res. spectrum \\
\citet{obs24} & Chandra & Carina Nebula & $-$ & spectrum \\
\citet{obs25} & XMM-Newton & colliding-wind stars & Mg XI \& XII, Si XIII & high res. spectrum \\
\citet{obs26} & Swift & C/2007 N3 (Lulin) & $-$ & image, spectrum \\
\citet{obs27} & Chandra, XMM & Pl. Nebula A30 & C VI & spectrum, low S/N \\
\citet{obs28} & Chandra & 5 comets & Mg, Si & spectrum \\
\citet{obs29} & Chandra & 103P/Hartley 2 & $-$ & image, spectrum \\
\citet{obs30} & Suzaku & Earth geocorona & $-$ & spectrum \\
\citet{roberts2015} & Suzaku & Cygus Loop SNR & O VII triplet & spectrum \\
\citet{obs32} & Chandra & 2 comets & $-$ & image, spectrum \\
\citet{obs33} & Suzaku & M82 & Ne X & spectrum \\
\citet{obs34} & XMM-Newton & sample of clusters & S XVI at 3.5~keV & spectrum \\
\citet{obs35} & Hubble & NGC~1275 AGN & Ne X, S XV & spectrum, 2$\sigma$ \\
\citet{gu2016a} & XMM-Newton & C/2000 WM1 & $-$ & high res. spectrum \\
\citet{obs37} & XMM-Newton & C/2000 WM1 & $-$ & high res. spectrum \\
\citet{obs38} & XMM-Newton & sample of clusters & O VIII & high res. spectrum \\
\citet{obs39} & XMM-Newton & Earth geocorona & $-$ & spectrum \\
\citet{obs40} & XMM-Newton & NGC 5548 AGN & N VII & high res. spectrum \\
\citet{obs41} & XMM-Newton & NGC 5548 AGN & O VI & high res. spectrum \\
\citet{obs42} & XMM-Newton & M51 & O VII, O VIII & high res. spectrum \\
\citet{walker2015} & Chandra & Perseus cluster & $-$ & spectrum \\
\citet{obs43} & Hitomi & Perseus cluster & S XVI & high res. spectrum \\
\citet{obs45} & XMM-Newton & Cygnus Loop SNR & $-$ & high res. spectrum \\
\citet{obs46} & RXTE & EXO 1745-248 burst & Ti, Cr, Fe, Co & spectrum \\
\citet{obs47} & Hitomi & Perseus cluster & S XVI, Fe XVII & high res. spectrum \\
\citet{obs48} & XMM-Newton & G296.1-0.5 SNR & $-$ & high res. spectrum \\
\citet{obs49} & XMM-Newton & N132D SNR & $-$ & low significance \\
\citet{obs50} & XMM-Newton & J0453.6-6829 SNR & O VII & high res. spectrum \\
\citet{obs51} & XMM-Newton & 1E 0102.2-7219 SNR & $-$ & high res. spectrum \\
\citet{obs52} & XMM-Newton & NGC~4636 & O VII & high res. spectrum \\
\citet{brb2007} & XMM-Newton & Jupiter & $-$ & high res. spectrum \\
\citet{obs54} & Chandra & SWCX & O VII, O VIII & spectrum \\
\citet{wulf2019} & XQC & SWCX & C, N, O & high res. spectrum \\
\citet{obs55} & XMM-Newton & M82 & N, O, Ne & high res. spectrum \\

\hline
\end{longtable}

\subsection{Solar system objects}

On 25 March 1996, comet C/1996 B2 (Hyakutake) had a close encounter with Earth at a distance of $\sim$ 0.1 AU. Rosat took the opportunity to search for X-ray signal from this unusual event. While astrnomers expected very low luminosity because comets are too cold to emit X-ray, the Rosat image showed a surprisingly bright source from a crescent-shaped region on the sunward side of comet \citep{obs1}. Follow-up studies of Rosat archival data proved that X-ray emission is common in comets. One year later after the Rosat discovery, \citet{cravens1997} solved the mystery: the X-ray emission originates from the charge exchange between highly charged heavy solar wind ions and neutral gas around the comet atmosphere. The spatial and spectral properties predicted by the charge exchange model has been found to be consistent with Rosat data of other comets, as well as the later Chandra and XMM-Newton observations with better spatial and spectral resolutions \citep{dennerl2010}.   

That the 1996 Rosat data established the charge exchange study of solar system objects may seem odd, as a number of earlier studies (e.g., \citealt{hh1962, gom1983}) already pointed out that charge exchange is non-negligible in the solar wind interaction. However, the early works tended to underestimate or overlook the charge exchange contribution in the X-ray band. The Rosat discovery greatly expanded the field, leading to a number of dedicated astrophysical observations, but also stimulating relevant laboratory measurements and theoretical calculations. A new focus of the laboratory and theoretical studies is to reproduce the X-ray observations, with highly-charged ions (C, N, O, Ne, etc) and cometary neutrals colliding at a speed of several hundred kilometer per second \citep{bei2003}.

\subsubsection{Comets}

\begin{figure}[bht]
\sidecaption
\includegraphics[width=\textwidth]{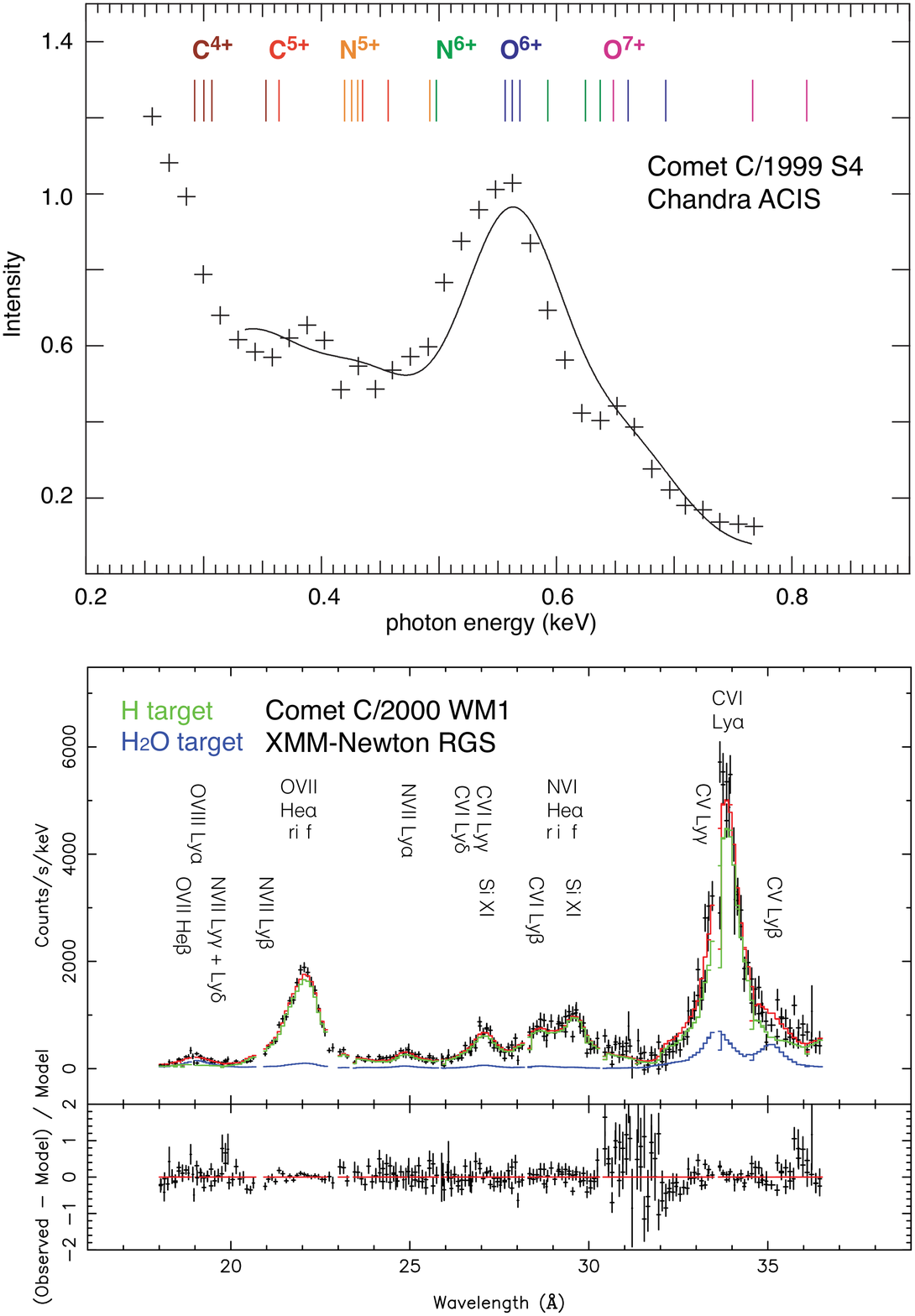}
\caption{(Upper) Comet C/1999 S4 {\it Chandra} CCD spectrum taken on 14 July 2000, compared with the model of C, N, O ions charge exchanging with CO$_2$ measured at EBIT \citep{exp51}. (Lower) Comet C/2000 WM1 {\it XMM-Newton} grating spectrum fit with the SPEX {\it cx} model taking into account the hydrogen and water targets. }
\label{fig:comet}       
\end{figure}

In the context of solar wind charge exchange, comets are often used as natural probes of the latitude- and time-dependent properties of the solar wind ions, which produce emission features that can be recognized in the X-ray spectra. The solar wind study with comets benefits from the facts that the comets approach the Sun in a much broader range of latitudes than the planets, and their occurrence in a proper distance range is quasi-constant on year basis. The representative X-ray observations of comets are summarized in Table~\ref{obs}. 

The solar winds are found to be largely bimodal at solar minimum: the fast wind with velocity $\sim 700$ km s$^{-1}$, characterized by low density and low ionization, launched probably from the open magnetic fields regions at high solar latitudes. The slow wind, blowing at $v \sim 400$ km s$^{-1}$ from the solar equatorial region with a closed magnetic field, is often found to be denser and more ionized. When the Sun moves towards its maximum, the fast and slow winds become more mixed up.

Line diagnostics with the cometary X-ray spectra provide useful constraints on the speed and the chemical composition of the solar wind (e.g., \citealt{obs11,obs14}). In addition, the X-ray morphology of comets has been used to infer the shape and the spatial structure of bow shocks \citep{weg2005}. Finally, \citet{obs37} derived a constraint on the neutral composition of the cometary atmosphere using a global fit to the {\it XMM-Newton} grating spectrum.

Two typical comet spectra are shown in Figure~\ref{fig:comet}. The {\it Chandra} observation of C/1999 S4 reveals a solar wind composition of C, N, and O in fully striped and hydrogen-like states. The {\it XMM-Newton} grating spectrum of C/2000 WM1 (linear) further resolves the strong individual C, N, and O lines, posing even tighter constraints on the solar wind abundance and even the chemistry of the cometary atmosphere. These data showcase the cometary X-ray as a powerful solar wind probe. However, most of the X-ray spectra are taken with a low CCD-like resolution of $\sim 100$ eV, making it impossible to study the solar wind dynamics using comets. A further advance in the X-ray study of comets requires observations using high-resolution X-ray spectrometers, in particular the microcalorimeters onboard XRISM \citep{tashiro2018} and Athena \citep{nandra2013} missions. 

\subsubsection{Planets}

\begin{figure}[bht]
\sidecaption
\includegraphics[width=\textwidth]{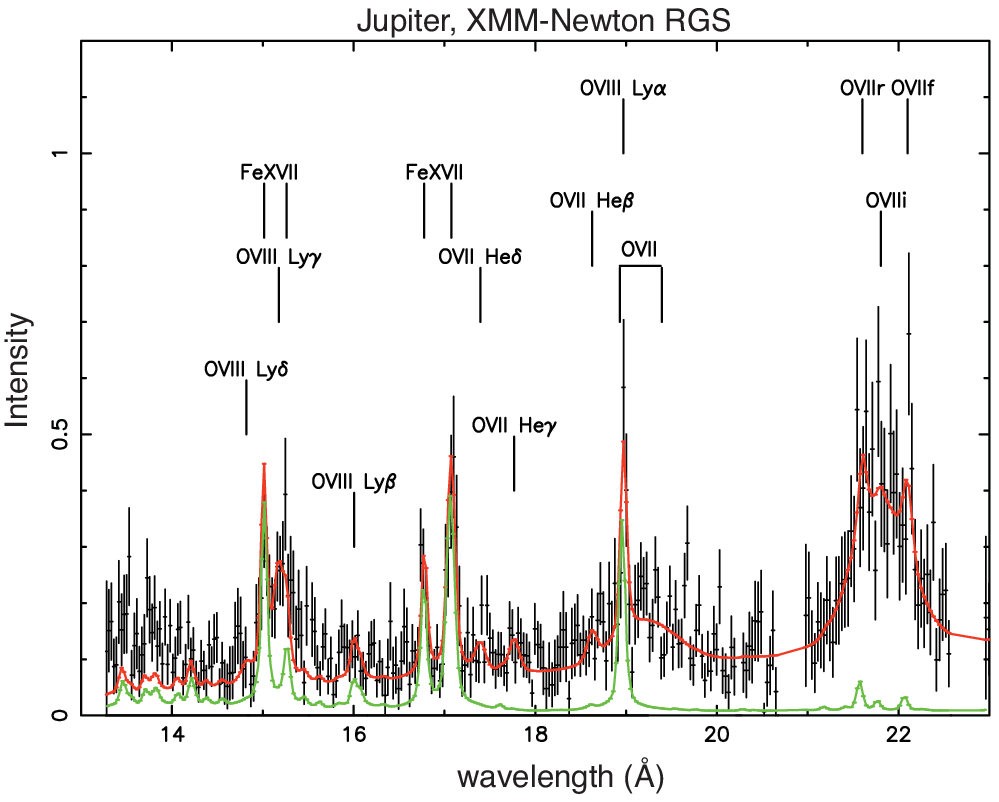}
\caption{The {\it XMM-Newton} grating spectrum of Jupiter. A phenomenological fit to the spectrum is plotted in red, and the disk component is shown in green. The emission from X-ray aurorae is the difference between the two. }
\label{fig:jupiter}       
\end{figure}

Jupiter is known as a bright X-ray source from the early Einstein observations \citep{metzger1983}. The origin of the X-ray has been settled just recently: the emission from the equatorial region is likely the scattering of solar X-ray in the dense planetary atmosphere, while the X-ray in the aurorae can be a mixture of two components, the bremsstrahlung radiation dominating $> 2$ keV by energetic electrons precipitating from the magnetosphere, and the charge exchange in soft X-ray between the in-falling ions and the atmosphere c neutrals \citep{gladstone2002, bhardwaj2006, br2004, bra2007, brb2007, br2008, kimura2016, dunn2017, numazawa2021}. The ion diagnostics using charge exchange provides an opportunity to understand where the ions come from: the X-ray spectra would be C- and O- rich if the solar wind is the main source of the ions, and S- and O is rich if the bulk population originates from the Io plasma torus falling into the Jovian magnetosphere.  

The {\it XMM-Newton} reflection grating spectrometer (RGS) spectrum of Jupiter in 2003 shown in Figure~\ref{fig:jupiter} show so-far the best example of the charge exchange diagnostics with X-ray spectroscopy of planets. It shows a clear separation of the charge exchange emission, mostly in oxygen lines, from the low-latitude disk component with Fe and O lines. The charge exchange lines are broadened by a velocity of $\sim 5000$ km s$^{-1}$, implying that we are witnessing a bulk of accelerated particles precipitating into the Jovian polar area.

Mars and Venus are possible charge exchange emitters, even though the detection was made at a lower significance than Jupiter \citep{obs9, obs13}. The scatter of solar X-rays in the planetary atmosphere is constantly bright in, even dominating, the overall soft X-ray band, making it challenging for charge exchange observation. No significant charge exchange signal has yet been detected in other non-Earth planets.

\subsubsection{Geocorona and heliosphere}

\begin{figure}[bht]
\sidecaption
\includegraphics[width=\textwidth]{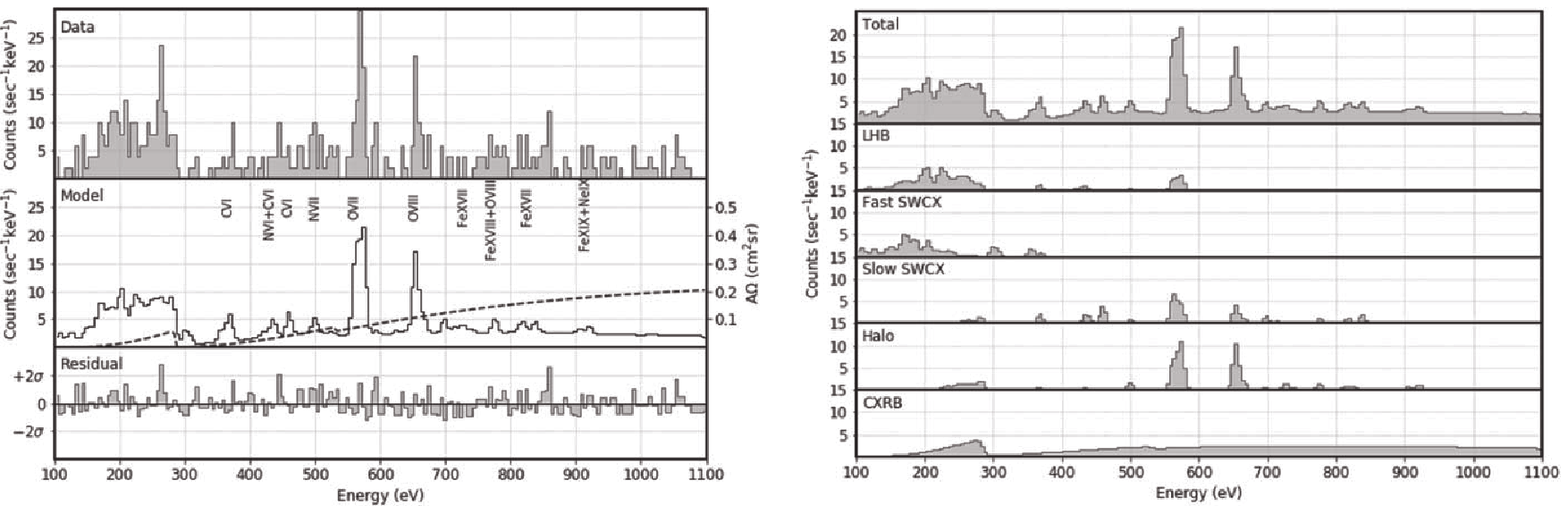}
\caption{(Left) X-ray quantum calorimeter spectrum of the diffuse soft X-ray background. The model is shown in the middle, and the residual is plotted at the bottom. The effective area of the detector is shown in the dashed line. (Right) The model is divided into components: LHB - local hot bubble; Fast/slow SWCX - fast/slow solar wind charge exchange; Halo - Milky way gas halo; and CXRB - cosmic X-ray background. }
\label{fig:swcx}       
\end{figure}

Charge exchange emission is generated when the highly ionized solar wind
interacts with neutral or near-neutral gas in the Earth's exosphere and the heliosphere. This component is commonly known as the solar wind charge exchange (SWCX), which can affect any X-ray observation so far. Therefore, it is treated as a troublesome background, consisting of multiple emission lines (e.g., C, N, and O) in the soft X-ray band. Concerning its temporal, spectral, and spatial variations, it is often challenging to fully eliminate the SWCX from the extrasolar sources.

{\it Chandra} observation of the dark moon has been considered direct evidence for the nearby SWCX since it shows charge exchange-type spectrum \citep{obs6}. Similarly, \citet{obs54} reported an enhancement of O VII and O VIII lines from SWCX in the {\it Chandra} observation of molecular cloud MBM~12, and \citet{obs12} observed a transient increase of SWCX flux by comparing several spectra of a {\it Suzaku} blank field in the direction of the north ecliptic pole. 

In a CCD resolution, the X-ray spectra from the geocoronal and heliospheric SWCX could sometimes resemble those from the local hot bubble around the solar system, making it difficult to isolate one from another in the spectrum. One solution is to separate them based on directional variation: the local hot bubble is isotropic while the SWCX might increase in the focusing cone of heliospheric helium caused by the inflow of interstellar medium. By scanning the sky through the helium cone using the diffuse X-rays from the local galaxy (DXL) mission, \citet{gale2014} separated the SWCX and the local hot bubble emission based on their spatial distributions. Their results showed that the SWCX emission contributes about 40\% of the 1/4 keV X-ray flux in the Galactic plane.

A second approach is to distinguish SWCX using high-resolution X-ray spectroscopy. The X-ray quantum calorimeter (XQC, $\sim 6$~eV FWHM), onboard the University of Wisconsin-Madison/Goddard Space Flight Center sounding rocket, has measured several times the diffuse X-ray background below 1~keV \citep{wulf2019}. As shown in Figure~\ref{fig:swcx}, the high-resolution spectroscopy allows the separation of the SWCX  from the 0.1~keV local hot bubble and the 0.2~keV Milky Way halo components. It could be used to further isolate the charge exchange from fast solar wind and slow wind components.

\subsection{Astrophysical objects}

While the charge exchange from the solar system objects is well established, the detection in extrasolar astrophysical objects remains somewhat controversial though evidence is accumulating in particular in the past decade. A general issue is that many claims are made with the CCDs on {\it Chandra}, {\it XMM-Newton}, and {\it Suzaku} with $50-100$~eV resolutions, or with dispersive gratings (e.g., RGS) of better but source-dependent resolutions. A revisit with non-dispersive high resolution X-ray spectrometers (e.g., XRISM and Athena) is needed to verify these features.

\subsubsection{Stars and supernova remnants }

\begin{figure}[bht]
\sidecaption
\includegraphics[scale=0.7]{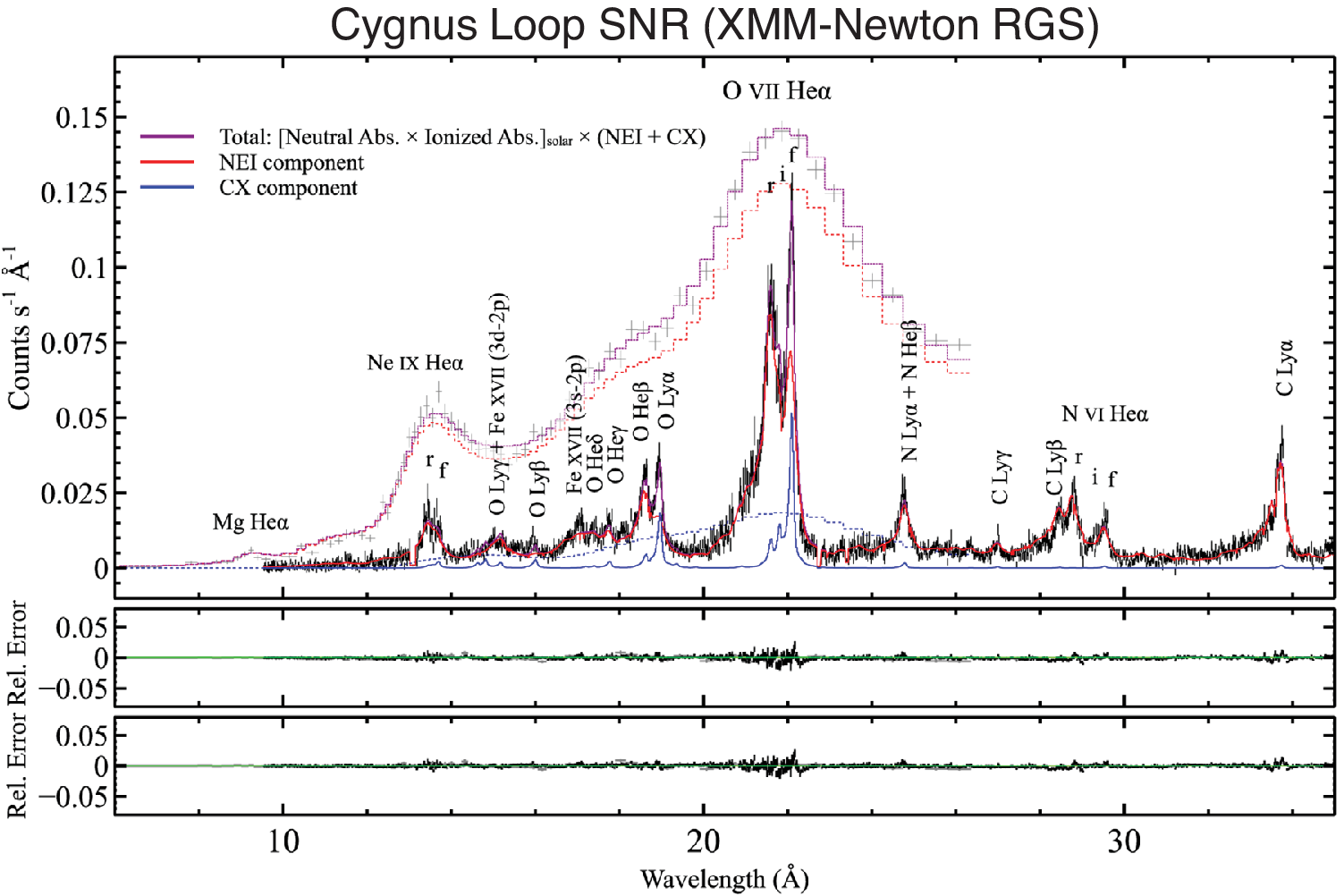}
\caption{{\it XMM-Newton} grating spectrum of a bright knot in the Cygnus Loop supernova remnant, fit with the charge exchange plus non-equilibrium collisional ionization model. The neutral and ionized absorbers are taken into account. }
\label{fig:snr}       
\end{figure}

\citet{obs25} discussed a scenario that colliding wind ejected from binary stars might lead to ion-ion collision and charge exchange. The ion-ion charge exchange can be realized only when the collisions occur at a large velocity ($\geq 2000$ km s$^{-1}$), which indeed resembles some of the observed terminal shock speeds of the colliding stellar winds. \citet{obs25} suggested that the charge exchange recombination might explain the low Mg XII/Mg XI ratios observed in systems including WR22, WR140, and $\zeta$ Pup.

The wind interaction condition might apply to other types of stellar objects. \citet{obs46} reported the detection of a 6.6 keV emission feature in the spectrum of a neutron star X-ray binary EXO 1745-248 40 hours after its superburst. It is a broad feature with an equivalent width of 4.3~keV, which is much larger than that of the standard ionized Fe line formed by the reflection in the disk. A spectral analysis with the RXTE data implied that the feature can be a set of charge exchange lines, likely emitted from the interaction of a failed wind launched by the superburst with the neutrals in the neutron star disk.

The interaction between supernova remnants and ambient neutral clouds might generate charge exchange emission. It is expected that this emission would be visible around the edge of the remnants, where interstellar neutrals pass through the shock front and collide with the post-shock hot ions. \citet{obs20} reported a detection of charge exchange emission at the outer shell of the Cygnus Loop, which shows an enhanced feature at the energy of highly excited O VII lines in the {\it Suzaku} spectrum \citep{roberts2015}. As plot in Figure~\ref{fig:snr}, \citet{obs45} took an {\it XMM-Newton} RGS spectrum of a bright X-ray spot on the outer shell, and showed that the charge exchange component, together with an ionized absorber, are indeed required to explain the forbidden-to-resonance line ratio of O VII. Similarly, the RGS spectra of G296.1-0.5 \citep{obs48}
and J0453.6-6829 \citep{obs49} show high forbidden-to-resonance ratios that can be explained by charge exchange.

The limited imaging capability of {\it XMM-Newton} RGS allows brief mapping of the ion-neutral interaction region in the supernova remnants. \citet{obs45} pioneered this method by differentiating the O VII forbidden line and resonance line monochromatic images, and discovered that the charge exchange lines can be emitted from a compact region where a dense neutral cloud is also found.

\subsubsection{Interstellar medium and star forming regions }

\begin{figure}[bht]
\sidecaption
\includegraphics[width=\textwidth]{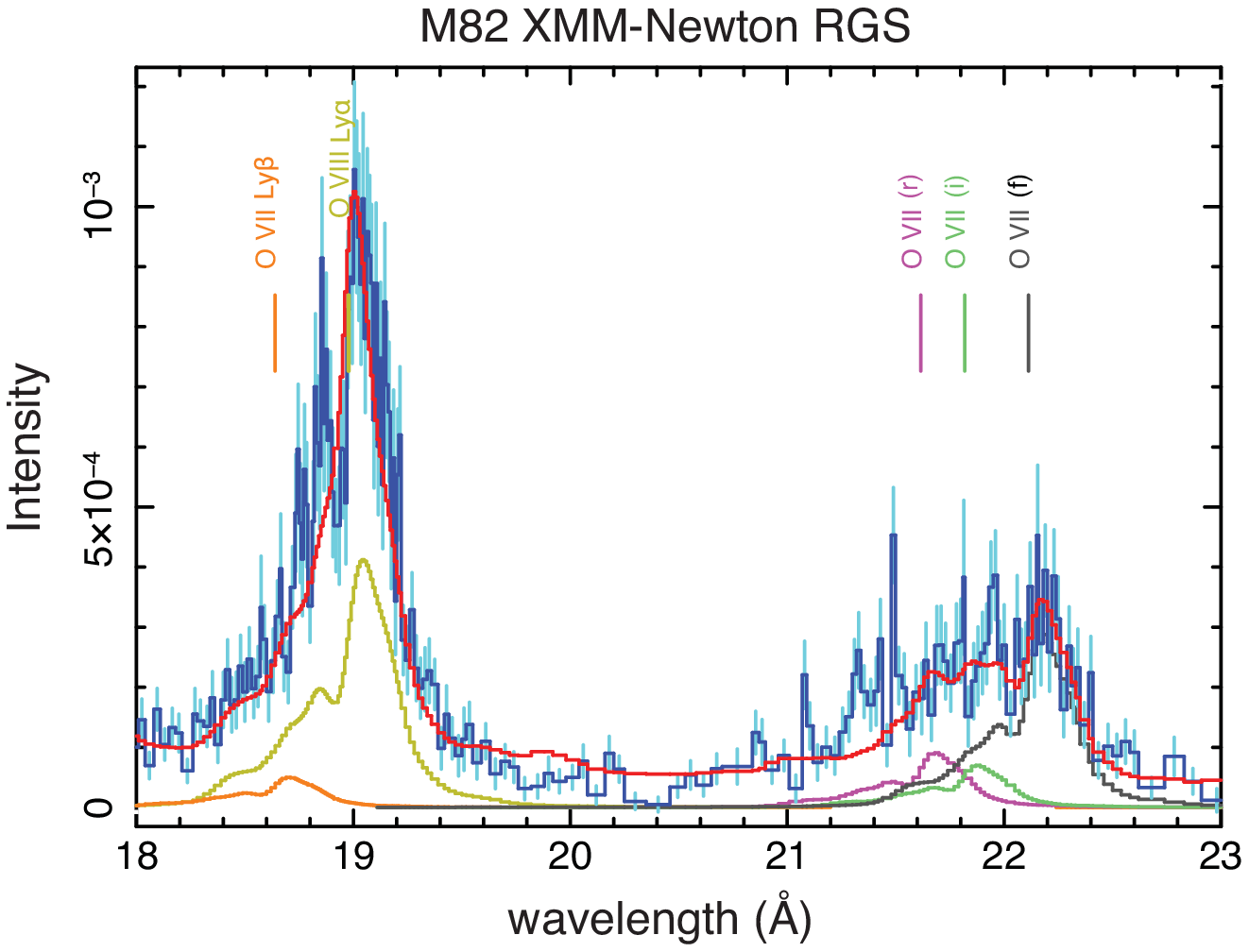}
\caption{Part of the {\it XMM-Newton} grating spectrum of the M82 star-forming galaxy. The charge exchange lines from O VII and O VIII are plotted in colors indicated by the label. }
\label{fig:m82}       
\end{figure}

The north polar spur is a large ridge of X-ray and radio-emitting gas in the northern sky. Similar to some of the supernova remnant spectra discussed above, the north polar spur also shows large forbidden-to-resonance ratios for He-like triplets of O and Ne \citep{miller2008, lallement2009}, which might potentially indicate a charge exchange origin. However, this scenario is far from conclusive since the forbidden-to-resonance ratio could also be explained by others atomic physics including an ionized absorber. As suggested in \citet{gu2016b}, the ionized absorption might be responsible for most of the observed line ratios while the charge exchange contributes no more than $10$\% to the forbidden lines.

The star-forming regions in normal or starburst galaxies are another locations where interactions between hot gas and cold clouds are expected. By modeling the {\it Chandra} spectrum of the Carina star-forming complex with multiple thermal components, \citet{obs24} reported several line-like excesses in the residual in $0.5-2.0$~keV. The distribution of these line residuals is spatially correlated with the cold star-forming filaments, where the gas-filament charge exchange might be responsible for these line emissions.

There have been claims for the detection of charge exchange X-ray in several starburst galaxies. M82 is the best-studied object in this category. The X-ray spectroscopic works using {\it XMM-Newton} and {\it Suzaku} data showed that the thermal model is insufficient in particular for the O VII and O VIII lines, while adding an additional charge exchange component could fix most of the problem \citep{tsuru2007,analli2008, obs21, obs22, obs33, obs55}. A recent study using the RGS spectrum is plotted in Figure~\ref{fig:m82}. \citet{obs23} presented a sample study of nine star-forming galaxies with {\it XMM-Newton} RGS, and showed that most of these galaxies are undergoing charge exchange as indicated by the forbidden-to-resonance line ratios. Therefore charge exchange can be a class property of star-forming regions and galaxies. Naturally, the ion-neutral interaction could occur between the starburst-driven outflows and the cold interstellar medium. 

In the quiescent galaxies, charge exchange might still occur at various locations including the interfaces between the nucleus-driven outflows and the circumnuclear neutral clouds. \citet{obs42} reported a possible detection of charge exchange component which contributes significantly to the N, O, and Ne lines, in particular the He-like forbidden lines, in the RGS spectrum of M51 nucleus. The interaction between the radio jet and the cold interstellar medium might be responsible for the emission. Yet another possibility is that these lines originate from photoionization by the AGN outbursts in the past.

\subsubsection{Active galactic nuclei }

\begin{figure}[bht]
\sidecaption
\includegraphics[width=\textwidth]{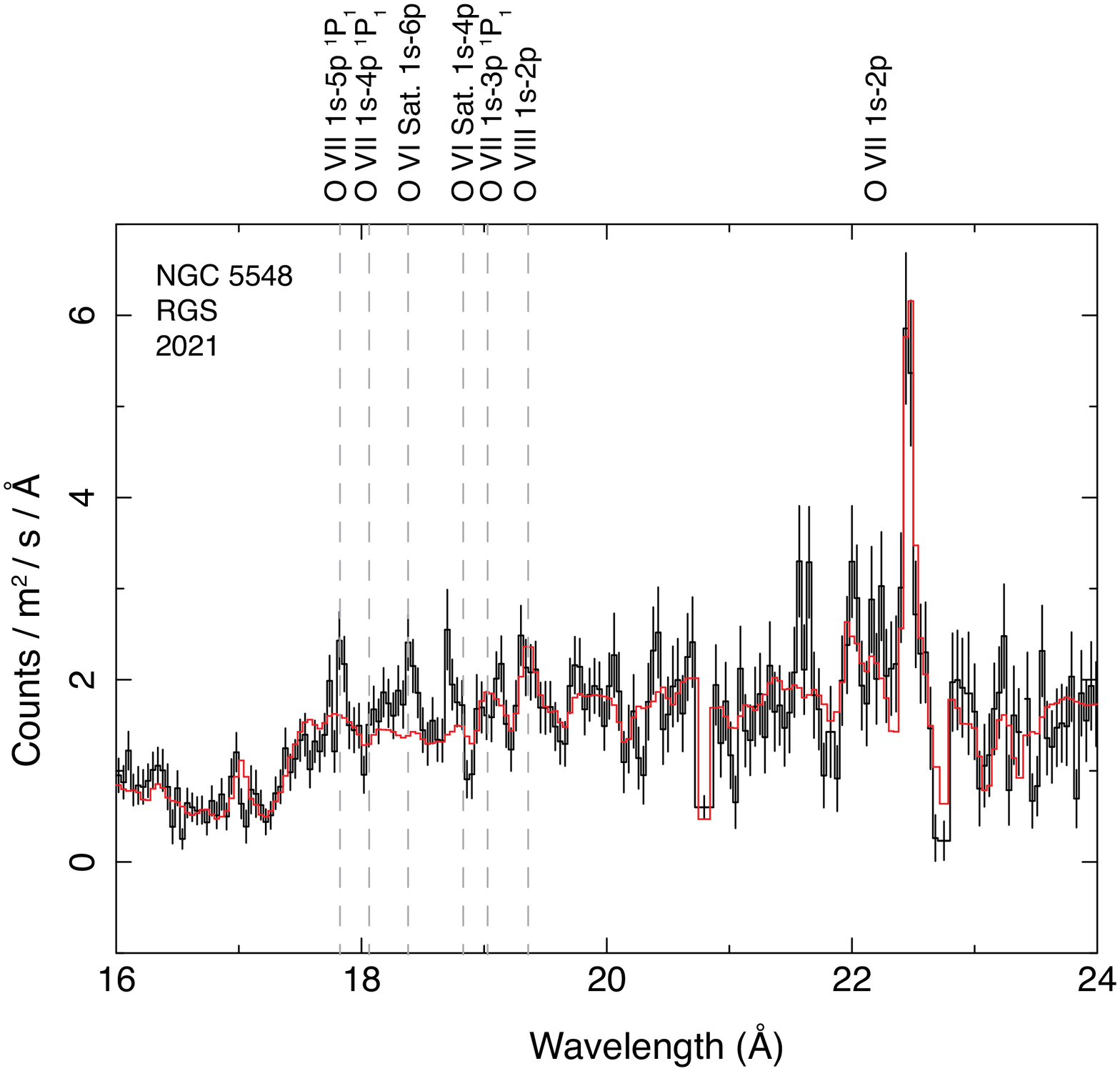}
\caption{{\it XMM-Newton} RGS spectrum of NGC~5548 AGN taken in January 2021 fit with the photoionization model. A set of excess features are seen around 18.4~{\AA} indicating the possible charge exchange emission. }
\label{fig:n5548}       
\end{figure}

AGNs host multiple gas inflows and outflows of a broad range of ionization states, making it another potential target of the charge exchange search. \citet{obs41} reported a detection of unidentified features at 18.4~{\AA} in the grating spectra of Seyfert I AGN NGC~5548. By stacking all observations, this feature is seen at $> 5\sigma$ significance taking into account the look-elsewhere effect. As indicated in Figure~\ref{fig:n5548}, the wavelength of the anomaly coincides with the high-$n$ transitions from He- and Li-like oxygen, therefore, it is likely a charge exchange line. The authors suggested that the interaction could happen within the same outflow by mixing the partially ionized and neutral layers, or between the outflow with the neutral close environment. Another report by \citet{obs35} showed that the Hubble STIS spectrum of the radio galaxy NGC~1275 can be modeled by including three weak lines at 1223.6~{\AA}, 1242.4~{\AA}, and 1244.0~{\AA}, each with a significance of $2-3 \sigma$. These features can be explained by a mixture of charge exchange between highly ionized hydrogen, neon, and
sulfur with neutral matter, indicating for an outflow with $v \sim 3400$ km s$^{-1}$.

\subsubsection{Clusters of galaxies}

\begin{figure}[bht]
\sidecaption
\includegraphics[width=\textwidth]{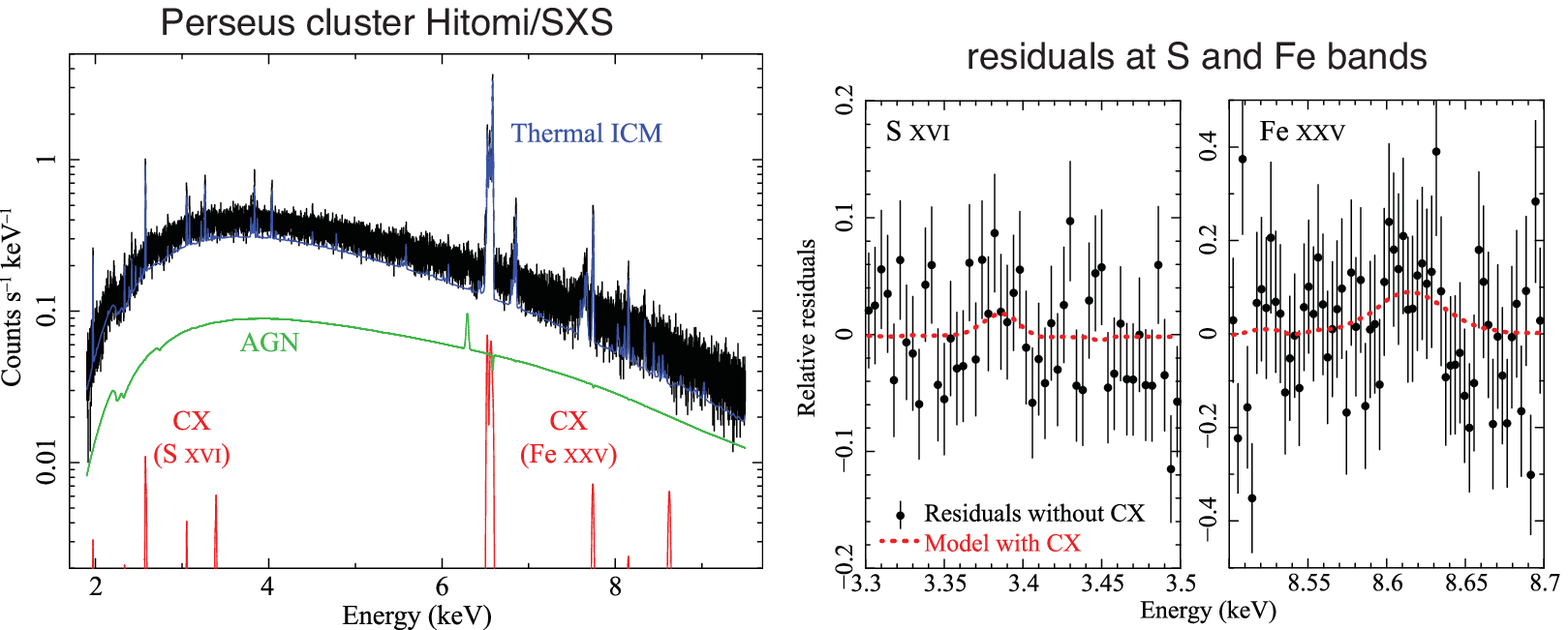}
\caption{(Left) {\it Hitomi} spectrum of the Perseus cluster fit with a sum of thermal, AGN, and charge exchange components. (Right) Residuals of the fit with the thermal plus AGN model. The red curve in each panel shows the model change by including the charge exchange component. }
\label{fig:perseus}       
\end{figure}

Though the hot plasmas are the dominating baryonic component in clusters of galaxies, cold gas clouds do exist, and they are often observed in and near the central galaxies, as well as in the wake of member galaxies during their infall to the center \citep{con2001}. These neutral structures are immersed in the giant pool of hot, highly ionized plasma, and their interfaces are candidates for charge exchange. \citet{fabian2011} and \citet{walker2015} suggested that the charge exchange can account for a part of the H$\alpha$ and soft X-ray emission from bright filaments in the Perseus cluster. 

During its brief lifetime, the micro-calorimeter onboard {\it Hitomi} observed the Perseus cluster with a $\sim 5$~eV resolution in the $2-10$~keV band. As shown in Figure~\ref{fig:perseus}, there are indeed hints for charge exchange in the {\it Hitomi} spectrum. The high-$n$ Rydberg transitions are found to be $1.6\sigma$ and $2.4\sigma$ significance for S XVI and Fe XXV. The detection remains challenging even for {\it Hitomi}, since the charge exchange in Perseus cluster is overwhelmed by the thermal emission by two orders of magnitude in flux. The S XVI charge exchange is in particular interesting since it nearly overlaps in energy with the mysterious 3.5~keV line detected in a large sample of clusters by \citet{bulbul2014} and \citet{boyarsky2014}. The 3.5~keV line was originally proposed as an evidence for the radiative decay of sterile neutrino, a theoretical form of dark matter. As shown in \citet{obs43}, the Hitomi spectrum of the Perseus cluster prefers the S XVI charge exchange scenario, although the data quality is insufficient to rule out the dark matter line. In addition, \citet{obs38} reported a 2.8$\sigma$ detection of a line feature at 14.82~{\AA}, which is likely a O VIII charge exchange line. The line flux measured with the RGS is in good agreement with the possible {\it Hitomi} detection.

\section{Ending remarks}
\label{sec:4}
The introduction of the charge exchange process to general X-ray astronomy was made in 1996 by the discovery of X-ray from a comet. It has been broadly understood now that, in parallel to the electron impact and photon induced X-ray, the ion-neutral interaction can also efficiently generate X-ray in astrophysical objects. In the past two decades after the cometary discovery, X-ray astronomers have spotted charge exchange-like X-rays in a variety of objects, ranging from the neighbour planets and heliosphere to the interstellar medium, galaxies, supermassive black holes, and clusters of galaxies on the cosmological scales. Charge exchange emission is no doubt the best tool for detecting an interface, because the normal electron-impact X-ray flux emitted from the physically-thin interface is likely much dimmer. It has the potential to change the research landscape on the interactions of (1) AGNs and the host galaxies; (2) galactic outflow and interstellar medium; (3) infalling galaxies and the cluster.

Charge exchange produces only line emission, making it a science case tailored for high-resolution spectroscopy. With low-resolution spectra it often remains challenging to disentangle charge exchange from a thermal or photoionized component. Once fully resolved, charge exchange lines provide diagnostics on (1) the collision speed between ions and neutrals; (2) the chemical composition and ionization state of the hot ions; (3) relation with the H$\alpha$-emitting region; and (4) the neutral species. Spatially-resolved spectroscopy can further constrain the interaction region.

New astrophysical discoveries could feedback to the theoretical and laboratory studies of charge exchange. Substantial work remains to remove the tension between the theoretical calculations of cross sections and the corresponding laboratory measurements. The Li-like and ion species with more than three electrons are still poorly covered in the current calculations and experiments. The $l$- and $S$- distributions of the electron capture are crucial for an accurate line model, but they remain less certain than the $n$- distribution. Finally, the multi-electron charge exchange remain to be addressed. A consistent and continuous effort will be required to ensure that the atomic data are ready for the next generation of high-resolution X-ray data to be obtained with XRISM and Athena.

\begin{acknowledgement}
SRON is supported financially by NWO, the Netherlands Organization for Scientific Research. C.S. acknowledges support from NASA under award number 80GSFC21M0002 and Max-Planck-Gesellschaft (MPG).
\end{acknowledgement}

\bibliographystyle{unsrt}


\end{document}